# Electric field non-linearity in very high frequency capacitive discharges at constant electron plasma frequency

Sarveshwar Sharma[1], Nishant Sirse[2], Animesh Kuley[3] and Miles M Turner[2]

[1]Institute for Plasma Research and HBNI, Gandhinagar – 382428, India
[2]School of Physical Sciences and NCPST, Dublin City University, Dublin 9, Ireland
[3]Department of Physics, Indian Institute of Science, Bangalore 560012, India

E-mail: nishudita1628@gmail.com

## Abstract

A self-consistent particle-in-cell simulation study is performed to investigate the effect of driving frequency on the electric field non-linearity, electron heating mechanism and *electron energy distribution function* (EEDF) in a low pressure symmetric capacitively coupled plasma (CCP) discharge at a constant electron plasma frequency. The driving frequency is varied from 27.12 MHz to 100 MHz for a discharge gap of 3.2 cm in argon at a gas pressure of 1 Pa. The simulation results provide insight of higher harmonic generations in a CCP system for a constant electron response time. The spatio-temporal evolution and spatial time-averaged electron heating is presented for different driving frequencies. The simulation results predict that the electric field non-linearity increases with a rise in driving frequency along with a concurrent increase in higher harmonic contents. In addition to the electron heating and cooling near to the sheath edge a positive <J.E> is observed in to the bulk plasma at higher driving frequencies. The EEDF illustrate enhancement in the population of mid-energy range electrons as driving frequency increases thereby changing the shape of EEDF from bi-Maxwellian to nearly Maxwellian. For the constant ion flux on the electrode surface, a decrease in the ion energy by more than half is observed with an increase in driving frequency.

# 1. Introduction

For several decades, capacitively coupled plasma (CCP) discharges excited at 13.56 MHz driving frequency are utilized for plasma processing applications, such as plasma enhanced chemical vapour depositions (PECVD) and reactive ion etching (RIE). It is known that the excitation of CCP's at frequencies above 30 MHz , i.e. in the VHF range (30-300 MHz), offers great benefits in comparison to 13.56 MHz. In particular, a higher plasma density can be achieved at lower discharge voltages along with a reduction in the ion energy (lower self-bias voltages) toward the surface, which suppresses the ion induced damage. The previous studies [1-3] shows that a higher deposition rate could be achieved (without affecting the film quality) at higher driving frequencies when compared to traditional 13.56 MHz CCP's. Similar process improvement have been obtained in the case of reactive ion etching where higher driving frequency CCP's showed enhanced etch selectivity to $SiO_2$ [4].

The benefits of increasing driving frequency in CCP's are related to a transition in the ion and electron heating dynamics, which significantly affect the plasma parameters including electron energy distribution function. At lower driving frequencies, i.e. in the range of few hundreds of kHz to few MHz, both ion and electrons responds to time varying electric field applied between the electrodes and therefore gain energy. As driving frequency increases to 13.56 MHz or even higher, only electrons respond instantaneously to time varying electric field, whereas, ion responds to time-averaged field. In a low pressure regime, the electron gain energy mostly through stochastic or collisionless heating process via interaction with high voltage oscillating sheaths [5-15]. This heating mechanism is known as α-heating mode. The energetic electrons produced during this heating process could be confined in the discharges if they hit opposite sheath during the expanding phase, and thus enhance the overall electron energy. This phenomenon is recognized as electron bounce resonance heating mechanism [16]. The presence of self-excited plasma series resonance can further enhance the stochastic heating [17-21]. As driving frequency increases, the energy gained by the electron from oscillating sheaths increases in comparison to energy gained by the ions. It is observed that the VHF driven CCP discharges are mostly sustained by the energetic electron beam generated from near to the sheath edge [21-23]. These electron beams can generate strong electric field transients in the bulk plasma causes heating of low energy electrons, which are mostly confined in the bulk plasma due to ambipolar field [23]. Multiple electron beams are also observed in low pressure CCP's excited in VHF regime [23]. It is proposed that the combination of discharge voltage and driving frequency can achieve an independent control of ion flux and ion energy [24].

At higher driving frequencies, the generation of higher harmonics in the CCP system is an another interesting feature. These higher harmonics are very efficient in terms of power deposition within a discharge system. In an earlier study by Miller et al [25], measured higher harmonics using magnetic probe in a CCP system for 3 different driving frequencies. Higher harmonic contents up to the 10$^{th}$ harmonic at 60 MHz driving frequency was observed in their study. Another work by Upadhyay et al [26] investigated electromagnetic wave phenomena in an axisymmetric CCP using numerical simulations. In the wave electric field, their study predict significant higher harmonic contribution up to the 20$^{th}$ harmonic of the fundamental driving frequency. Both studies revealed that the enhanced local power depositions by the higher harmonics is one of the factors responsible for plasma non-uniformity at VHF driven CCP's. A simulation study performed by Wilczek et al [27] for voltage and current driven CCP's showed that the higher harmonics in the rf current is observed in the case of voltage driven CCP's. However, no such harmonics are present in voltage at the electrode when a current driven CCP is simulated. The higher harmonics contents also shown to enhance the excitation of plasma series resonance, which can lead to a substantial increase in the plasma density. Our recent simulation studies [28] predict that these higher harmonics are the function of discharge voltage. As discharge voltage increases, the higher harmonics and therefore non-linearity in the CCP increases. In view of practical applications, it is highly important to explore the high harmonic generation and their effect on plasma parameters to better understand the power coupling mechanism in VHF CCP systems.

The present work investigate the higher harmonic generation, electron heating mechanism and electron energy distribution function versus driving frequency at constant electron plasma frequency. The study is performed using self-consistent particle-in-cell/Monte Carlo collision simulation technique. Most of the driving frequencies chosen in the present study are similar to one used in commercial processing reactors. The constant electron plasma frequency case is considered here, which will provide more insight of higher harmonics generation since then their generation will be mostly dependent on the driving frequency and sheath properties. The current simulation study is highly relevant for processing applications since it simulate the ion energy for different driving frequencies at constant ion flux. The simulation results, carried out for a constant electron plasma frequency, presents important information about the energy distribution of the electrons i.e. the electron energy distribution function, which influence the plasma kinetic processes.

The paper is organized as follows. In section 2, we provide a description of the simulation scheme, initial condition set-up and discharge parameters considered in this work.

The physical interpretation and discussion of the simulation results are presented in section 3. Finally, a brief conclusion and summary remarks are given in section 4.

## 2. Simulation scheme and discharge parameters

A geometrically symmetric capacitively coupled plasma (CCP) discharge driven by radio-frequency (RF) is simulated by using the 1D-3V electrostatic particle-in-cell (PIC)/ Monte-Carlo Collisions (MCC) simulation technique [29,30]. We have considered voltage driven case for our present study. This code is developed at Dublin City University (DCU) by Prof. Miles Turner and it is well-tested PIC code, which is used in several research papers published in different reputed journals [23, 28, 31-36]. A thorough description about the simulation approach is described is literature [37-38]. It is assumed in simulation that both the electrodes have infinite dimension and are parallel to each other. So effectively the spatial direction is taken perpendicular to both the electrodes. The electrodes are perfectly absorbing for both the electrons and ions and for simplicity secondary electron emission is not considered in present research work. We have used argon plasma and all important particle-particle reactions like ion-neutral (elastic, inelastic and charge exchange) and electron-neutral (elastic, inelastic and ionization etc.) are considered in all sets of simulation. Other complex plasma reactions for e.g. multi-step ionization, metastable pooling, partial de-excitation, super elastic collisions and further de-excitation are not included in simulation for the sake of simplicity. The source of important reactions and their cross-sections are well-tested and reported in literature [34, 39-40]. The reactions responsible for metastable production (i.e. A* and A**) are included in the set of plasma reactions however we did not track them in output diagnostics. The choice of grid size and time step is in such a way that it can resolve the Debye length ($\lambda_{De}$) and electron plasma frequency ($\omega_{pe}$) respectively. These conditions are sufficient to fulfill stability and accuracy conditions of PIC technique. The gap between the electrodes is chosen 3.2 cm, which is divided in 512 number of grids for all cases. The number of particles per cell is 100 which is reasonable choice for PIC simulations. The neutral gas is 1 Pa (7.5 mTorr) and is distributed uniformly in background and its temperature is same as ion temperature i.e. 300 K. One of the electrode (at x=0 cm) is driven by following voltage waveform

$$V_{rf}(t) = V_0 \sin(2\pi f_{rf} t + \phi)$$

while the other electrode (at x=L i.e. electrode gap) is grounded one. The simulation is run for few thousand RF cycles to achieve steady state solutions.

## 3. Results and Discussions

This section present simulation results describing effect of driving frequency on the electric field transients, electron heating and electron energy distribution function (EEDF) at constant electron plasma frequency ($f_{pe} = \sqrt{ne^2/m\varepsilon_0}/2\pi$, where *n* is electron density, *e* is electronic charge, *m* is mass of electron and $\varepsilon_0$ is vacuum permittivity). Figure 1, shows the time-averaged electron and ion density profiles for different driving frequencies from 27.12 MHz to 100 MHz. The different values of driving frequencies considered here such as 27.12 MHz, 60 MHz and 100 MHz are similar to one used in the processing reactors. At 13.56 MHz, steady state is not observed for the given discharge parameters and therefore not shown here for the comparison. At each driving frequency, the discharge voltage (corresponding values are shown in the figure 1) is varied in order to maintain constant time-averaged plasma density in the system. As predicted, a decrease in the value of discharge voltage is observed at higher driving frequencies for maintaining same plasma density. This is attributed to a decrease in sheath impedance with a rise in driving frequency, which allows higher current to flow through the discharge and therefore higher plasma density is obtained at lower discharge voltage [41]. The electron and ion density is nearly same and highest at the centre of the discharge, whereas, it decreases towards the electrodes. The central plasma density is ~2.5×10$^{15}$ m$^{-3}$ for all cases and the corresponding electron plasma frequency is approximately 450 MHz. The plasma density at the sheath edge is also approximately constant with a value of (1.2-1.4)×10$^{15}$ m$^{-3}$, and the sheath width is decreasing from ~7 mm to ~5 mm for 27.12 MHz to 100 MHz respectively. The method for estimating sheath width is described later in the manuscript.

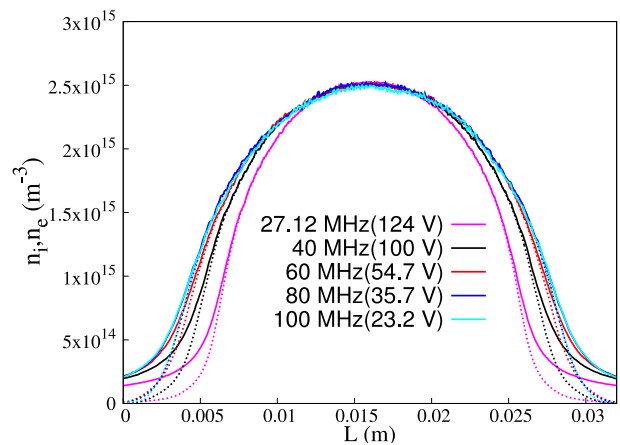

**Figure 1.** Time-averaged electron and ion density profiles for different driving frequencies.

Figure 2 (a), (b) and (c) shows the spatio-temporal evolution of electric field within the discharge system at 27.12 MHz, 60 MHz and 100 MHz respectively for last 2 RF cycles. As shown in figure 2, the electric field is mostly confined in the sheath region near to the electrodes. At 27.12 MHz, the electric field is nearly zero in the bulk plasma. As driving frequency increases to 60 MHz, the presence of strong electric field transients is observed in the bulk plasma. These electric field transients are extending up to the opposite sheath. At 100 MHz driving frequency, these electric field transients becomes more energetic, interacting with the opposite sheath and strongly perturbing the instantaneous sheath edge position. The presence of the strong electric field transients are due to the high energy electrons generating from near to the sheath edge. In the present case, the electron plasma frequency is approximately constant and therefore the energy imparted to the electron by oscillating sheaths will depend on the driving frequency. As driving frequency increases, the sheath expansion period decreases, which in turn increases the sheath velocity. Additionally, the high frequency oscillation on the instantaneous sheath edge position, such as at 100 MHz, will further increases the conjugate sheath edge velocity. An increase in the overall sheath edge velocity increases the energy gained by electrons from oscillating sheaths and therefore they are able to reach up to the opposite sheath. Being higher in energy, these energetic electrons can generate non-linear effects in the plasma. It could also heat bulk electrons effectively through non-linear mechanisms, which can lead to a change in the shape of EEDF. The EEDF is described later in the manuscript.

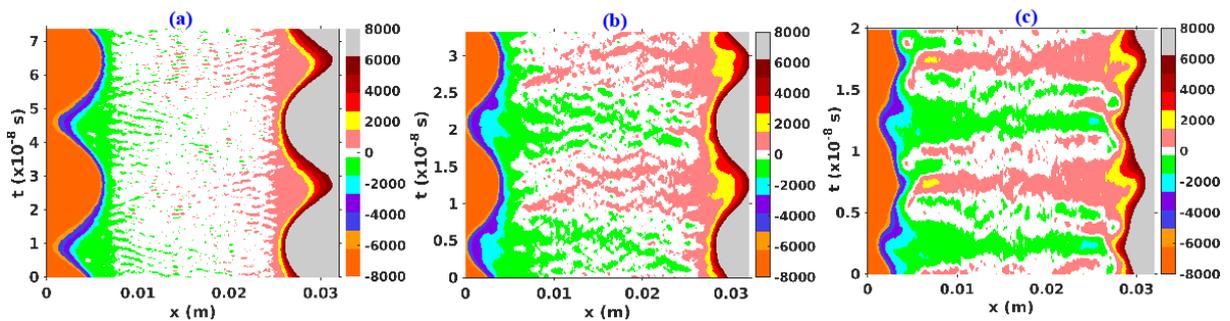

**Figure 2.** Spatio-temporal evolution of electric field in the discharge for a) 27.12 MHz, b) 60 MHz and c) 100 MHz driving frequency.

The non-linearity in the discharge can be revealed by transforming electric field from time domain to frequency domain using Fast Fourier Transform (FFT). Figure 3, shows the temporal evolution of electric field and its FFT at the centre of the discharge for the same driving frequencies as plotted in figure 2. As displayed in figure 3 (i, b), at 27.12 MHz the amplitude of the fundamental driving frequency is lowest (~5%). The 17th harmonic here is nearly same as the oscillations near to electron plasma frequency i.e. at 450 MHz. As driving frequency increases to 60 MHz, the fundamental contribution increases to approximately 18% (figure 3 (ii, b)). The electric field near to electron plasma frequency, 7th and 8th harmonics, are still appeared with a slight increase in their amplitude. In addition, higher frequency components above $f_{pe}$ i.e. 13th and 15th harmonics of the driving frequency are noticed in the electric field. Increasing driving frequency up to 100 MHz further increases the amplitude of the fundamental harmonic, which is approximately 25% of the total electric field (figure 3 (iii, b)). The electric field around electron plasma frequency increases slightly and higher harmonic contents up to 1.5 GHz is observed. These results verify that as driving frequency increases for constant electron plasma frequency, the electric field amplitude at fundamental driving frequency increases. The non-linearity (higher harmonic contents) in the discharge also increases with driving frequency. These higher harmonics are very effective in power depositions and can effectively heat the bulk plasma electrons.

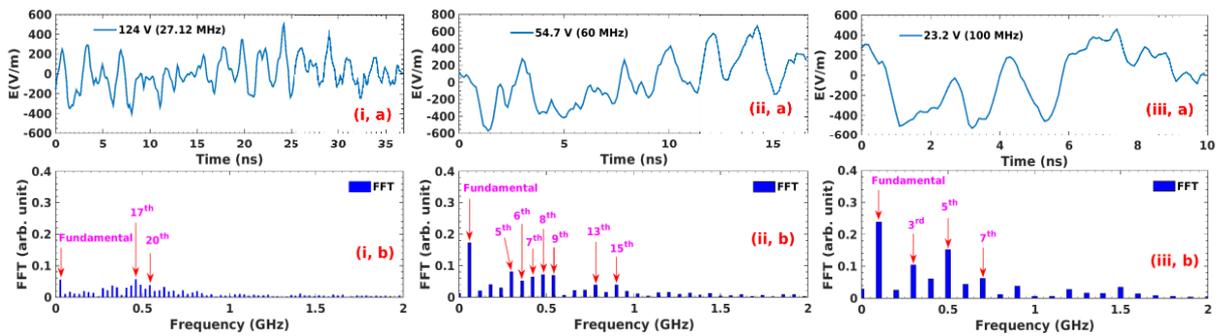

**Figure 3.** Electric field and its FFT at the centre of the discharge for a) 27.12 MHz, b) 60 MHz and c) 100 MHz driving frequency.

We next turn to electron heating mechanism in the discharge system at different driving frequencies. Figure 4 shows the spatio-temporal evolution of electron heating in the discharge at a) 27.12 MHz, b) 60 MHz and c) 100 MHz driving frequency and for last 2 RF cycles. The results shows that the electrons are gaining energy mostly during the phase of sheath expansion and losing energy during collapsing phase of the sheath. The gain and loss of the electron energy is maximum near to the sheath edge. In addition, further electron heating and cooling is observed in the bulk plasma at 60 MHz driving frequency. At 100 MHz driving frequency, the overall electron heating in the bulk plasma is more in comparison to electron cooling. To predict the total electron energy gain and loss, the time-averaged electron heating is plotted in figure 5 for different driving frequency. As shown in figure 5, positive <J.E> is observed near to the sheath edge, which shows that the electron heating is maximum at this location. As driving frequency increases, the electron heating near to the sheath edge first increases, up to 40 MHz, and then remains constant. At 60 MHz, a slight decrease in the electron heating is observed when compared to 40 MHz. In the bulk plasma, negative <J.E> i.e. electron cooling is observed up to 40 MHz driving frequency. Above 40 MHz, positive <J.E> is noticed in the bulk plasma and its region (area and magnitude of heating) increases with a further rise in the driving frequency up to 100 MHz. Along with positive <J.E> in the bulk plasma above 60 MHz, a strong negative heating is observed near to the sheath edge position. This region of negative heating region is responsible for driving electric field transients in the bulk plasma as shown in figure 2. The area under negative <J.E> increases with a rise in driving frequency, which is in consistent with the enhancement in positive <J.E> region inside the bulk plasma. Above results suggest that the 60 MHz is the transition frequency below which, heating occur mostly near to the sheath edge and electron cooling in the bulk plasma, whereas, above 60 MHz positive electron heating is also appeared in the bulk plasma due to energetic electrons.

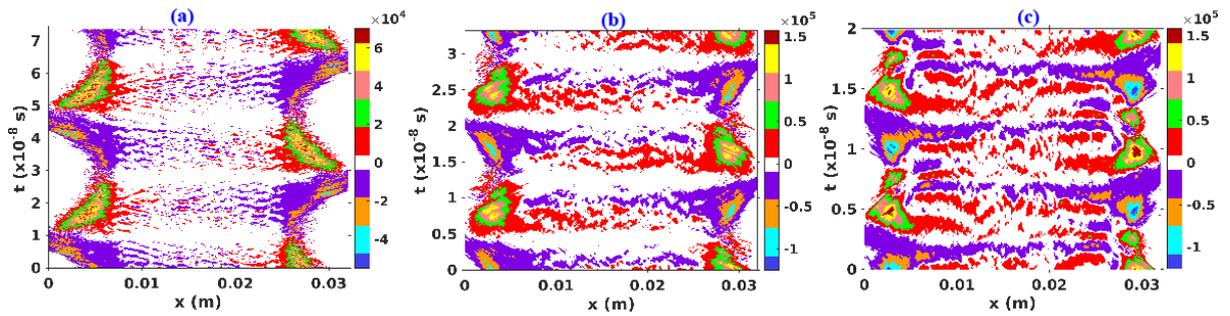

**Figure 4.** Spatio-temporal evolution of electron heating in the discharge for a) 27.12 MHz, b) 60 MHz and c) 100 MHz driving frequency.

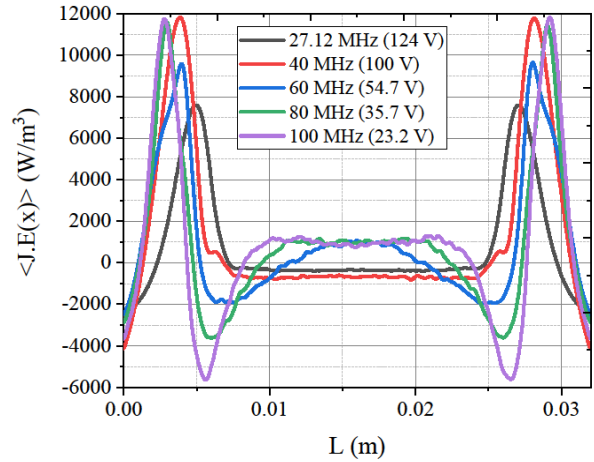

**Figure 5.** Time-averaged electron heating in the discharge at different driving frequencies and for constant electron plasma frequency.

Finally, the effect of driving frequency and transients on the electron energy distribution function (EEDF) is investigated. Figure 6 shows the EEDF at the centre of the discharge for 3 driving frequencies; 27.12 MHz, 60 MHz and 100 MHz. As displayed in figure 6, at 27.12 MHz the EEDF is bi-Maxwellian with a large population of low energy electrons and remaining electrons are preset in the high energy tail. As driving frequency increases to 60 MHz and 100 MHz, the shape of EEDF turns in to nearly Maxwellian i.e. single electron temperature. The population of the low energy electron decreases and the electron population having energy between 1 eV to 20 eV increases. The transformation of the EEDF is attributed to transfer of energy from electric field transients to the bulk electrons through non-linear interaction. As shown in figure 2 and 3 earlier that at 27.12 MHz driving frequency no electric field is present in the bulk plasma. The low energy electrons are mostly confined in the bulk plasma due to ambipolar field and therefore unable to gain energy from the high voltage oscillating sheaths, whereas, high energy electrons interact with the sheath and gain energy. This turn the shape of EEDF to bi-Maxwellian. As driving frequency increases to 60 MHz and 100 MHz, the electric field in the bulk plasma at fundamental driving frequency increases and higher harmonics appears. This is due to high energy electrons originating from near to the expanding phase of the sheath. These energetic electrons then redistributes their energy with the bulk electrons through non-linear mechanism and therefore causing increase in their energy. These bulk electrons after gaining energy diffuses in to the higher energy region of EEDF and turn the shape of EEDF to nearly Maxwellian. This transformation is highly beneficial for generating reactive species by excitation and dissociation mechanisms in the plasma when molecular gases are employed for plasma production/processing. It is further observed that the mid energy range electrons in the case of 100 MHz is slightly lower than 60 MHz. It is mainly due to the loss in the energy of electrons when they interact with opposite sheath. As shown in figure 2 (c), at 100 MHz driving frequency, the electric field transients are extending up to the opposite sheath. It is observed that some of the energetic electrons are hitting to the opposite sheath at the collapsing phase of the sheath around 10 ns and therefore lose their energy. The analysis of the ion energy distribution function (IEDF) on the surface showed that the ion flux is constant (~ $2\times10^{18}$ $m^{-2}s^{-1}$) for different driving frequencies considered in the present study. However, the energy of the ions decreases by more than half with driving frequency i.e. from ~75 eV at 27.12 MHz to ~35 eV at 100 MHz. Therefore, it can be concluded that by keeping the ion flux constant, the ion bombardment energy can be lowered by increasing the driving frequency along with the efficient production of reactive gas phase chemistry, which is mostly driven by the shape of EEDF.

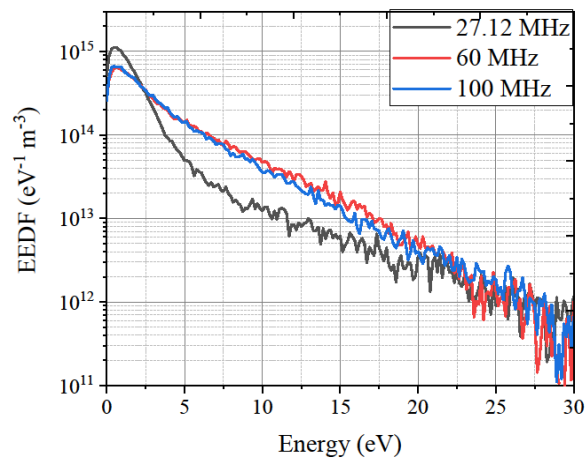

**Figure 6.** Electron energy distribution function at the centre of the discharge for 27.12 MHz, 60 MHz and 100 MHz driving frequency.

## 4. Summary and Conclusions

A low pressure symmetric capacitively coupled plasma discharge is studied for different driving frequencies using self-consistent particle-in-cell simulation technique. In particular, the effect of driving frequency on the electric field non-linearity, electron heating mechanisms and the electron energy distribution function (EEDF) in the discharge is investigated for a constant electron plasma frequency. The simulation results reveals that at higher driving frequency the discharge system becomes highly non-linear and the higher harmonics are generated. At 27.12 MHz, the plasma bulk is quasi-neutral, however at 60 MHz and 100 MHz driving frequency, electric field transients are observed inside the bulk plasma. Along with an increase in the amplitude of fundamental electric field with driving frequency, the presence of higher harmonics contents up to 1.5 GHz (15$^{th}$ harmonic) is noticed in the bulk plasma at 100 MHz driving frequency. The electron heating is mostly localized near to the sheath edge and first increases with driving frequency and then remains constant. A positive <J.E> is observed in the bulk plasma at higher driving frequency. The time-averaged <J.E> represent negative heating i.e. electron cooling close to the sheath edge, which is the energy source for driving electric field transients into the bulk plasma. At 100 MHz, the negative <J.E> region is maximum and corresponding electric field transients are strongest, reaching up to the opposite sheath and modifying the instantaneous sheath edge position. A non-linear interaction between energetic electrons produced from near to the sheath and bulk electrons increasing their energy and turning the shape of EEDF from bi-Maxwellian to nearly Maxwellian. The corresponding electron population between 1-20 eV increases drastically. The flux of ion on the surface is reported to be constant with a decrease in their energy by more than half from 27.12 MHz to 100 MHz. The simulation results conclude that the higher plasma density at lower discharge voltage and higher driving frequency is mainly due to significant presence of higher harmonic contents. At a constant ion flux, the higher driving frequency drive EEDF with large population of mid energy range electrons, which has a great benefit for producing for reactive species when a molecular gas is employed for plasma processing.

**Acknowledgement:** Dr A Kuley is supported by Board of Research in Nuclear Sciences (BRNS Sanctioned No. 39/14/05/2018-BRNS), Science and Engineering Research Board EMEQ program (SERB Sanctioned No. EEQ/2017/000164).